\documentclass[times, twocolumn]{aastex62}
\usepackage{bm}
\usepackage{hyperref}

\graphicspath{{./}{figures/}}
\shorttitle{Dragonfly 44 in MOND and MOG}
\shortauthors{Haghi et al.}

\begin{document}
\label{firstpage}
%\pagerange{\pageref{firstpage}--\pageref{lastpage}}

\title{The Star Formation History and Dynamics of the Ultra-diffuse Galaxy Dragonfly 44 in MOND and MOG}
\correspondingauthor{Hosein Haghi}
\email{haghi@iasbs.ac.ir}

\author[0000-0002-0786-7307]{Hosein Haghi}
\affil{Department of Physics, Institute for Advanced Studies in Basic Sciences (IASBS), PO Box 11365-9161, Zanjan, Iran}

\author{Vahid Amiri}
\affiliation{Department of Physics, Institute for Advanced Studies in Basic Sciences (IASBS), PO Box 11365-9161, Zanjan, Iran}

\author{Akram Hasani Zonoozi}
\affiliation{Department of Physics, Institute for Advanced Studies in Basic Sciences (IASBS), PO Box 11365-9161, Zanjan, Iran}
\affiliation{Helmholtz-Institut f\"ur Strahlen-und Kernphysik (HISKP), Universit\"at Bonn, Nussallee 14-16, D-53115 Bonn, Germany}

\author{Indranil Banik}
\affiliation{Helmholtz-Institut f\"ur Strahlen-und Kernphysik (HISKP), Universit\"at Bonn, Nussallee 14-16, D-53115 Bonn, Germany}

\author{Pavel Kroupa}
\affiliation{Helmholtz-Institut f\"ur Strahlen-und Kernphysik (HISKP), Universit\"at Bonn, Nussallee 14-16, D-53115 Bonn, Germany}
\affiliation{Charles University in Prague, Faculty of Mathematics and Physics, Astronomical Institute, V Hole\v{s}ovi\v{c}k\'ach 2, CZ-180 00, Praha 8, Czech Republic}

\author{Moritz Haslbauer}
\affiliation{Helmholtz-Institut f\"ur Strahlen-und Kernphysik (HISKP), Universit\"at Bonn, Nussallee 14-16, D-53115 Bonn, Germany}

%==============================================
\begin{abstract}

The observed line-of-sight velocity dispersion $\sigma_{los}$ of the ultra diffuse galaxy Dragonfly 44 (DF44) requires a Newtonian dynamical mass-to-light ratio of $M_{dyn}/L_I=26^{+7}_{-6}$ Solar units. This is well outside the acceptable limits of our stellar population synthesis (SPS) models, which we construct using the integrated galactic initial mass function (IGIMF) theory. Assuming DF44 is in isolation and using Jeans analysis, we calculate $\sigma_{los}$ profiles of DF44 in Milgromian dynamics (MOND) and modified gravity (MOG) theories without invoking dark matter. Comparing with the observed kinematics, the best-fitting MOND model has $M_{dyn}/L_I = 3.6_{-1.2}^{+1.6}$ and a constant orbital anisotropy of $\beta=-0.5_{-1.6}^{+0.4}$. In MOG, we first fix its two theoretical parameters $\alpha$ and $\mu$ based on previous fits to the observed rotation curve data of The HI Nearby Galaxy Survey (THINGS). The DF44 $\sigma_{los}$ profile is best fit with $M_{dyn}/L_I=7.4_{-1.4}^{+1.5}$, larger than plausible SPS values. MOG produces a $\sigma_{los}$ profile for DF44 with acceptable $M_{dyn}/L_I$ and isotropic orbits if $\alpha$ and $\mu$ are allowed to vary. MOND with the canonical $a_0$ can explain DF44 at the $2.40 \sigma$ confidence level (1.66\%) if considering both its observed kinematics and typical star formation histories in an IGIMF context. However, MOG is ruled out at $5.49 \sigma$ ($P$-value of $4.07 \times 10^{-8}$) if its free parameters are fixed at the highest values consistent with THINGS data. %Last sentence altered by IB.

\end{abstract}


%For Hosein: the DF44 article should mention the colour is not too strong a constraint since the SFR ended long ago

%the likely smaller size in the past means the dynamical time would have been less
%the large size now suggests a short SFR to enable this growth and that the object is not spherical, so there may be some additional systematic uncertainty in our spherical dynamical modeling.

%the next email will contain the pages with my comments. Please include also the following reference on p.4:
%MOND as arising from the quantum vacuum:
%Smolin, 2017:
%https://ui.adsabs.harvard.edu/abs/2017PhRvD..96h3523S/abstract
%and on p.1, please include Famaey et al. 2018:
%https://ui.adsabs.harvard.edu/abs/2018MNRAS.480..473F/abstract

%Please write at the end of Sec.1: "Throughout this contribution we differentiate between the dynamical and stellar mass-to-light ratio in the photometric $I-$band, $M_{dyn}/L_I$ and $M_*/L_I$, respectively. The latter contains all stellar remnants. If Newtonian gravitation without dark matter were correct, the $M_{dyn}/L_I=M/L_I$. If Newtonian gravitation with dark matter in the galaxy were to be correct, then $M_{dyn}/L_I > M_*/L_I$. If MOND is correct (taking into account the external field effect), then $M_{dyn}/L_I = M_*/L_I$ always. These statements hold for the case that tidal field effects are not important."

\section{Introduction} \label{sec:intro}

Ultra diffuse galaxies \citep[UDGs,][]{vanDokkum15} are characterised by an extremely low surface brightness of $\mu_{(g, 0)}< 24$ mag arcsec$^{-2}$, comparable to those of dwarf galaxies. But unlike dwarfs, UDGs have extended stellar distributions, sometimes with a size comparable to that of the Milky Way ($\approx 4$ kpc).

\cite{vanDokkum+18a} and \cite{vD19_DF4} previously reported the discovery of two `dark matter-free' UDGs, NGC 1052-DF2 (DF2) and NGC 1052-DF4 (DF4), with line-of-sight velocity dispersions of $\sigma_{los}=8.5^{+2.2}_{-3.1}$ km/s \citep{Danieli_2019} and $\sigma_{los}= 4.2^{+4.4}_{-2.2}$ km/s \citep{vanDokkum+19}, respectively. In both cases, the upper limit to the halo mass is $M_{200} \la 10^8 M_{\odot}$ \citep{vanDokkum+18a, vD19_DF4, Wasserman18}. They argue that the apparent lack of dark matter in  DF2 and DF4 is in contrast to Dragonfly 44 (DF44), which is gravitationally dominated by dark matter \citep{vanDokkum16}. This large apparent difference in dark matter content is surprising as DF2, DF4 and DF44  have a very similar stellar mass and morphology.  %are all rich in globular clusters and

%They claimed that the predicted velocity dispersion of DF44 in MOND is lower than the observed value challenging the alternatives to dark matter theories such as MOND. 

UDGs are also interesting systems for testing classical and alternative gravity theories such as  Milgrom's modified Newtonian dynamics \citep[MOND,][]{Milgrom83} or Moffat's modified gravity \citep[MOG,][]{Moffat05}. UDGs generally experience internal accelerations below the characteristic MOND acceleration of $a_0= 1.2 \times 10^{-10} \, m/s^2$. Therefore, MOND predicts a different $\sigma_{los}$ than Newtonian dynamics in the absence of dark matter. In \citet{Kroupa_2018} and \citet{Haghi19}, we calculated $\sigma_{los}$ for the UDGs DF2 and DF4 using a new analytic formulation and fully self-consistent live $N$-body models in MOND. We showed that the velocity dispersion calculated by taking into account the external field effect (EFE) from possible host galaxies leads to  $\sigma_{los}$ being well consistent with the observed values in both cases. The same conclusions are reached independently by \citet{Famaey18}.

%could significantly reduce the MOND-predicted $\sigma_{los}$ in both cases. The reduction could plausibly be sufficient to explain the observed values, even though these are much less than the MOND predictions without the EFE.

%For these UDGs, the external acceleration due to the host galaxy and the internal acceleration due to the stars themselves are significantly below the critical limit of $a_0$ such that the expected velocity dispersions in the case of MOND exceed those expected in the classical Newtonian case by up to a factor of 3.

The velocity dispersion profile of UDGs can also be modeled using the appropriate Jeans equation, from which one can derive the statistical properties of the velocity distribution of stars \citep{BT08}. This calculation requires knowledge of the gravitational potential and the stellar density distribution, in addition to some assumptions about the shape of the velocity distribution.%function of stars in position and velocity determines the number of stars in a given region of space, and 

Taking $\sigma_{los}$ to be a function of radius alone for eight MW dwarf spheroidal galaxies (dSphs) and using spherical Jeans analysis, \citet{Angus08} showed that most of them have a MOND dynamical $M_{dyn}/L$ ratio compatible with stellar population synthesis (SPS) models. Using Jeans analysis, \citet{Haghi2016} studied the internal dynamics of eight MW dSphs in the framework of the MOG theory. They showed that when letting the two MOG parameters $\alpha$ and $\mu$ (Sec. \ref{MOG_basics}) vary on a case by case basis, the best-fitted $M_{dyn}/L$ ratios for almost all dSphs are comparable with the SPS values. However, this eliminates the predictability of the theory.

In this contribution, we focus on DF44, a UDG in the Coma Cluster \citep{vanDokkum16, DiCintio17}. Recently, \citet{vanDokkum+19} obtained a revised mean line-of-sight velocity dispersion within its half-light radius of $\sigma_{los}= 33$ km/s and thus an estimated halo mass of $M_{200} = 10^{11} - 10^{12} M_{\odot}$. DF44 has a rising $\sigma_{los}$ profile, from $\sigma_{los} = 26 \pm 4$ km/s at $R = 0.2$ kpc to $\sigma_{los} = 41 \pm 8$ km/s at $R = 5.1$ kpc, with no observed signs of rotation. They showed that this profile can only be fit with a standard $NFW$ halo if the velocity distribution has a strong tangential anisotropy ($\beta=-0.8^{+0.4}_{-0.5}$). A good fit also results from a dark matter halo with a relatively flat density profile \citep[e.g.][]{DiCintio14} and no orbital anisotropy ($\beta= -0.1^{+0.2}_{-0.3}$). \citet{vanDokkum+19} also calculated the mass profile of DF44, finding a Newtonian dynamical mass-to-light ratio in the photometric $I-$band of $M_{dyn} /L_I = 26^{+7}_{-6}$ Solar units within the effective radius of $\approx 3$ kpc. The reported $M_{dyn} /L_I$ is similar to other UDGs but $\approx 6 \times$ higher than normal galaxies of the same luminosity.

Using then available data, \citet{Hodson_2017} suggested that DF44 poses a problem for MOND, leading them to seek further modifications to the theory. The deep-MOND dynamical mass $\propto{\sigma^4_{los}}$, meaning that underestimated measurement errors on $\sigma_{los}$ can lead one to prematurely conclude against this highly non-linear theory.

\citet{Bilek_2019}  revisited the question of whether MOND can adequately explain the internal kinematics of DF44, taking into account the latest $\sigma_{los}$ measurements. Assuming isotropy of the velocity dispersion tensor and  $M_{dyn}/L_I=1.3$, they found that MOND matches the observed $\sigma_{los}$ in the central regions. The agreement is poorer in the outer regions, but the MOND prediction is still within 2$\sigma$ of the observations.

We use spatially resolved stellar kinematic data of DF44 to test MOND and MOG by solving the Jeans equation and determining the expected $\sigma_{los}$ profile. Since its stars experience an orbital acceleration $<a_0$, there will be clear water between the Newtonian and Milgromian predictions. We will show that the dynamics of DF44 can be explained in the MOND alternative to dark matter using an acceptable stellar $M/L_I$ ratio without making further ad hoc modifications to MOND. The difference between our conclusions and those of \citet{Hodson_2017} arise mainly from a subsequent downwards revision to $\sigma_{los}$ \citep{vanDokkum+19}. %IB: deleted mention of how MOG works fine without ad hoc modifications, since changing alpha and mu when the data doesn't agree with MOG could be considered ad hoc.

%Here we computed Jeans models for this
%galaxy in the framework of MOND. They found that an isotropic model reproduces the
%observed velocity dispersion near the center without any ad hoc
%tuning of free parameters. For the outer parts, the modeled velocity dispersion deviates from the %measured data points but it is
%still within the two sigma uncertainty limit of the measurement.
%While the modeled velocity dispersion profile is almost flat, the
%observed one is steeply rising in the outer parts.

Throughout this contribution we differentiate between the dynamical and stellar mass-to-light ratio in the photometric $I-$band, $M_{dyn}/L_I$ and $M_*/L_I$, respectively, with the latter including stellar remnants. The gravity theory used to calculate $M_{dyn}/L_I$ should be clear from the context. If Newtonian gravitation without dark matter were correct, then the Newtonian $M_{dyn}/L_I=M/L_I$. If Newtonian gravitation with dark matter in the galaxy were to be correct, then $M_{dyn}/L_I > M/L_I$. If MOND is correct (taking into account the external field effect), then the MOND $M_{dyn}/L_I = M_*/L_I$ always. These statements hold for the case that tidal effects are not important.

Our paper is organized as follows $-$ based on photometry of DF44 and our understanding of stars, the stellar population synthesis (SPS) prediction for its $M_*/L_I$ ratio is presented in Sec. \ref{SPS}. After briefly reviewing the basics of the MOND and MOG theories and introducing the Jeans equation (Sec. \ref{Basics}), we present our results (Sec. \ref{Results}). We then discuss our results in Sec. \ref{Discussion} and present our conclusions in Sec. \ref{Conclusion}.

\section{SPS prediction for the $M_*/L_I$ ratio in the IGIMF context} \label{SPS}

%In this section, we find the conditions under which SPS models can provide a $M_*/L_I$ ratio consistent with what we inferred from our dynamical modeling of DF44 in MOND (Sec. \ref{MOND_fit}). %Altered by IB, see below.

We begin by discussing our SPS models, which we use to provide a gravity model-independent expectation for the $M_*/L_I$ ratio of DF44. With the correct gravity theory, this must be consistent with what we infer from our dynamical modeling (Sec. \ref{MOND_fit}).

\subsection{The galaxy-wide IMF}

The stellar $M_*/L_I$ ratio depends on the age and metallicity of the stellar population and the stellar initial mass function (IMF), which in turn depends on the metallicity and star formation rate (SFR,  \citealt{Kroupa13, Yan+17, Jerabkova18, Zonoozi19}). In galactic environment studies, the stellar IMF is usually assumed to be invariant. But there are observational indications that the galaxy-wide IMF (gwIMF) may depend on the star formation environment (cloud density and metallicity), becoming top-heavy under extreme starburst conditions \citep{Dab09, Dab10, Dab12, Marks12, Banerjee12, Kroupa13, Schneider18, Kalari18, Jerabkova17}. The data suggest that with increasing embedded cluster metallicity and decreasing density, the IMF becomes less top-heavy. Changes to the IMF have also been proposed to explain $M_*/L$ ratios estimated through integrated light analysis of globular clusters (GCs) in M31, which show an inverse trend with metallicity \citep{Zonoozi16, Haghi17}. Another argument supporting a systematic variation of the IMF is the fraction of low-mass X-ray binaries in Virgo GCs and ultra compact dwarf galaxies \citep{Dab12}.

To quantify the gwIMF, \citet{KW03} formulated the integrated galaxy IMF (IGIMF) theory by assuming that all stars form in embedded clusters \citep{LL03, Kroupa05, Megeath16} and adding the IMFs of all clusters which form in a star formation epoch.  Based on the IGIMF theory, the gwIMF is top-light (deficit of massive stars) in low mass galaxies because they are expected to have a low SFR \citep{Ubeda07, Lee09, Watts2018, Yan+17}. The gwIMF is predicted to be more top-heavy in massive galaxies with a high SFR, as is observed \citep{ HG08, Lee09, Meurer09, Habergham10, Gun11, Zhang18, Hopkins18}. 

%There are also competing models for the deficiency of high-mass stars in low-density regions based on the stochasticity of a universal IMF in a study of deep $H_{\alpha}$ observations of the outskirts of M83 \citep{Koda_2012}.
There are also competing models based on the stochasticity of a universal gwIMF for the deficiency of high-mass stars in low-density regions, as suggested by deep $H_{\alpha}$ observations of M83's outskirts \citep{Koda_2012}. The stochastic models however face the challenge of needing to account for the systematic shift with increasing SFR of the observationally deduced gwIMF being top-light for dwarf disk and top-heavy for massive disk galaxies \citep{Lee09, Gun11}. Old, dormant galaxies also indicate significant gwIMF variations at the low-mass end $-$ elliptical galaxies may be dominated by very low mass stars \citep{vanDokkum10, vanDokkum11, Conroy17}, while ultra-faint dwarf galaxies may have a deficit of low mass stars when compared to the canonical stellar population \citep{Geha13, Gennaro+18}. Here, we study the influence of the gwIMF and star formation history (SFH) on the stellar $M_*/L_I$ ratio of DF44.

\subsection{The expected $M_*/L_I$ ratio in the IGIMF context}

\begin{figure}
\includegraphics[width=80mm, height= 70mm]{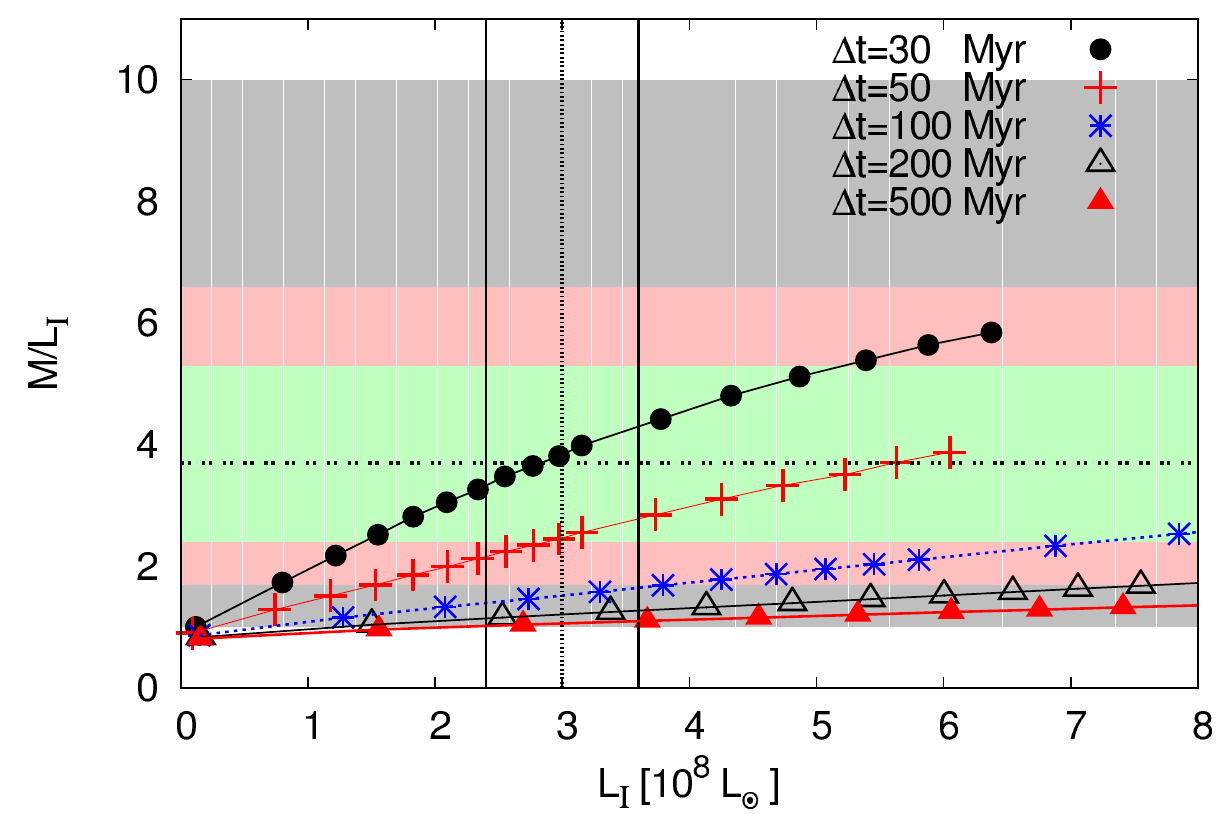}
\includegraphics[width=80mm, height= 70mm]{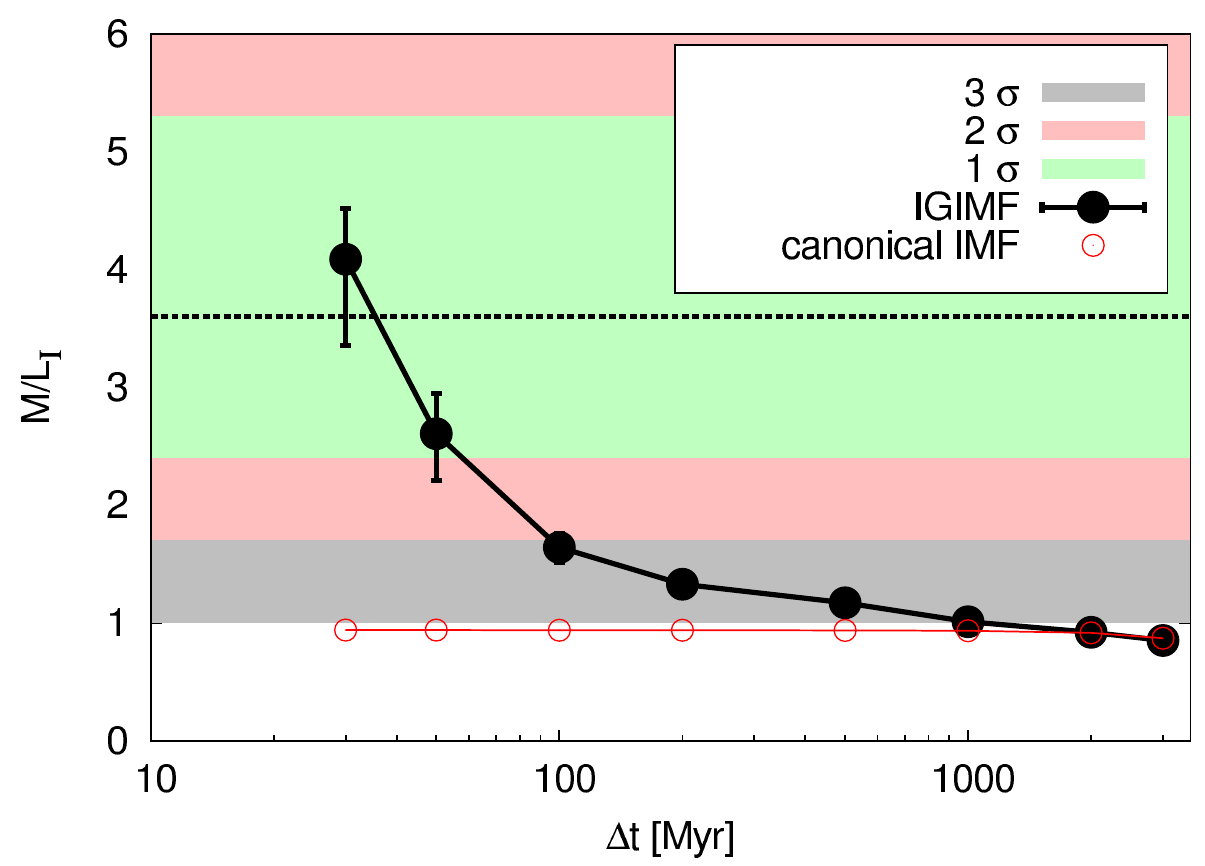}
\caption{\emph{Top}: $M_*/L_I$ versus the total $I$-band luminosity of a dwarf galaxy in the IGIMF context for different star formation durations of $\Delta t=$ 30, 50, 100, 200 and 500 Myr. The observed DF44 luminosity of $3 \times 10^8 L_{I\odot}$ is indicated by a vertical dotted line, with 20\% lower and higher values shown with solid vertical lines to its left and right, respectively \citep[section 6.2 of][]{vanDokkum+19}. The green, pink and grey horizontal bands show the 1, 2 and $3\sigma$ ranges, respectively, of the inferred $M_{dyn}/L_I$ in MOND, with the best-fitting value of 3.6 shown as a dotted line (Fig. \ref{f1}). \emph{Bottom}: $M_*/L_I$ ratio as a function of the star formation timescale $\Delta t$ for the observed $L_I$ of DF44 $\pm$ its $1\sigma$ uncertainty, obtained by interpolation in $L_I$ for the IGIMF (black circles) and invariant canonical (red circles) gwIMF. In the latter case, the error bars are smaller than the points $-$ this is also true for the IGIMF when $\Delta t \ga 0.2$ Gyr. We also ran the 1 Gyr and 3 Gyr models with an exponential SFH, but do not show the results as $M_*/L_I$ differs by $\la 0.05$ from the corresponding model with a constant SFH.}
\label{M_L_DF44}
% \end{center}
\end{figure}

Based on the canonical gwIMF \footnote{The `canonical gwIMF' or `canonical IMF', is the standard invariant two-part power-law IMF with Salpeter index of $\alpha_2=2.3$ for stellar mass $m> 0.5 M_{\odot}$ and $\alpha_1=1.3$ for smaller stellar masses.}, the $M_*/L_I$ ratio of a 10 Gyr old DF44-like galaxy is about 1 in Solar units. Assuming a bottom-heavy or top-heavy gwIMF, the $M_*/L_I$ ratio would increase owing to the higher fraction of low mass stars or stellar remnants, respectively. 

Here, we use the IGIMF theory to determine DF44's $M_*/L_I$ ratio. In the IGIMF context, the gwIMF would be top-heavy if the SFR $>1 M_{\odot}/yr$ \citep{Yan+17, Jerabkova18}. We assume a constant galaxy-wide SFR over the duration of star formation, $\Delta t$. Assuming an age of 10 Gyr for DF44 \citep{vanDokkum+19}, we calculate its $M_*/L_I$ ratio for different values of its total mass and $\Delta t$. The SFR increases with decreasing $\Delta t$, since $SFR=M_{tot}/\Delta t$, where $M_{tot}$ is the total mass of all stars were formed in the galaxy.

Fig. \ref{M_L_DF44} shows the calculated present-day $M_*/L_I$ ratio versus the total $I$-band luminosity for different $\Delta t$. The total luminosity of DF44 is $\left(3.0 \pm 0.6 \right) \times 10^8 L_{I\odot}$, which we show as vertical lines. The bottom panel of Fig. \ref{M_L_DF44} shows the $M_*/L_I$ ratio as a function of the star formation timescale $\Delta t$. Decreasing $\Delta t$ leads to a higher present-day $M_*/L_I$ ratio. This is because decreasing $\Delta t$ increases the SFR which, in the context of the IGIMF, implies a more top-heavy IMF and hence a higher mass-to-light ratio as well as a higher average stellar age. Assuming that $\Delta t > 30$ Myr, the $M_*/L_I$ ratio must be in the range $1-4 $.

However, the expected $M_*/L_I$ is almost independent of $\Delta t$ for the canonical IMF since the galaxy is much older than $\Delta t$ in the range considered (red line in Fig. \ref{M_L_DF44}). The minor decrease of the expected $M_*/L_I$ for the canonical IMF at $\Delta t \geq 2$ Gyr is due to the lower number of remnants formed for a longer duration of star formation.

\section{Modeling the velocity dispersion in MOG and MOND} \label{Basics}

%After briefly introducing the MOND and MOG theories, we use the Jeans equation to find the $\sigma_{los}$ profile in a spherically symmetric gravitational field. 

\subsection{Modified Gravity (MOG)} \label{MOG_basics}

In the weak field approximation, one can obtain the exact static spherically symmetric solution of the MOG field equation for a point-like mass by perturbing the fields around Minkowski space-time for an arbitrary distribution of non-relativistic matter \citep{Moffat13}. In this regime, the MOG gravitational potential $\Phi(\bm{x})$ and gravitational acceleration $a(\bm{x})=-\bm\nabla\Phi(\bm{x})$ are found using the principle of superposition as follows (see their equation 32):
\begin{equation}
\label{e6}
\Phi(\bm{x})\!=\!-G \!\! \left[\!\int\!\!\frac{\rho(\bm{x'})}{\mid\!\bm{x}\!-\!\bm{x'}\!\mid}(1\!+\!\alpha\!-\!\alpha e^{-\mu\mid\bm{x}-\bm{x'}\mid})d^3x' \!\right] \, ,
\end{equation}
%\begin{eqnarray} 	\label{e7}
%a(\bm{x})=-G\int\frac{\rho(\bm{x'})(\bm{x}-\bm{x'})}{\mid\bm{x}-\bm{x'}\mid}\,\,\,\,\,\,\,\,\,\,\,\,\,\,\,\,\,\,\,\,\,\,\,\,\,\,\,\,\,\,\,\,\,\,\,\,\,\,\,\,\,\,\,\,\,\,\,\,\nonumber\\
%\Big[1+\alpha -\alpha e^ {-\mu \mid \bm{x}- \bm{x'} \mid}(1+\mu\mid\bm{x}-\bm{x'}\mid)\Big]d^3x'.
%\end{eqnarray}
%These parameters determine the coupling strength of the vector field to baryonic matter and the range of the fifth force. 
Here, $G$ is the Newtonian gravitational constant, $\alpha$ is determined by the coupling strength of the fifth force vector, $\Phi_\mu$, to the baryonic matter and $\mu$ is the range of this force. The parameters $\alpha$ and $\mu$ are the free parameters of the theory that should be universal parameters fixed by observations \citep{Moffat_2009}. The resulting MOG acceleration from a point-like mass $M$ in the weak field regime contains a Yukawa-type force added to the Newtonian acceleration \citep{Moffat06}:
\begin{equation}
\label{e8}
a_{MOG}(\bm{r})=\frac{GM}{r^2} \lbrace{1+\alpha[1-e^{-\mu r}(1+\mu r)]\rbrace}.
\end{equation}
%IB deleted this from the first sentence: and $M$ is the total baryonic mass within the radius $\left| \bm{x} \right|$. Definition of $M$ added to last sentence.

The gravitational potential/field of an extended self-gravitating spherically symmetric system in MOG has been derived based on the point mass potential \citep{Moffat13, Roshan_2014}. 

By rotation curve analysis of the THe HI Nearby Galaxy Survey (THINGS) catalog of galaxies with a wide range of luminosities (from $L_B=7\times 10^7 - 5\times 10^{10} L_{B\odot}$), \citet{Moffat13} have shown that $\alpha$ and $\mu$ are universal parameters with the values $\alpha_{RC}=13$ and $\mu_{RC} = 0.15 kpc^{-1}$, respectively. They then applied the effective MOG potential with these fixed universal parameters to a sample of spiral galaxies in the Ursa Major catalog of galaxies and obtained good fits to galaxy rotation curve data. However, \citet{Haghi2016} showed that for almost all dSphs of the MW, the best-fitting values of $\alpha$ and $\mu$ are larger than $\alpha_{RC}$ and $\mu_{RC}$. They concluded that these parameters are not really universal constants in MOG as theoretically predicted \citep{Green_2018}, but instead take different values for different classes of objects. In Sec. \ref{Section_consistency}, we will confirm this conclusion again by showing that MOG with the universal parameters $\alpha_{RC}$ and $\mu_{RC}$ requires an unacceptable $M_{dyn}/L_I$ ratio for DF44.

%to fit MOG to the observed data of velocity dispersion as a function of radius
Since the total luminosity of DF44 ($L_I = 3\times 10^8 L_{I\odot}$) is well within the luminosity range of the dwarf galaxies in their sample, in model MOG1 (see Sec. \ref{Results}), we fix the MOG parameters to the upper limits obtained from the RC analysis of \citet{Moffat13}.

\subsection{Milgromian dynamics (MOND)}
\label{MOND_basics}

In the non-relativistic version of MOND interpreted as modified gravity \citep[comprehensively reviewed in][]{Famaey12}, the gravitational acceleration $g$ in an isolated spherically symmetric system is related to the Newtonian gravity $g_N$ by \citep{Milgrom83, Bekenstein_1984}
\begin{equation}
    g_N = g \mu \left( \frac{g}{a_0} \right) \, ,
\end{equation}
where Milgrom's constant $a_0 = 1.2 \times 10^{-10} \, m/s^2$ is the transition acceleration of the theory below which Newtonian dynamics breaks down \citep{Begeman91, Famaey07} and $\mu (x)$ is an interpolation function which is very close to $x$ when $x \ll 1$ but saturates at $1$ for $x \gg 1$. Different types of MOND interpolating functions have been used in the literature. The most common families of functions were reviewed in \cite{Famaey12}. The transition around $a_0$ can be interpreted to be due to the quantum vacuum \citep{Milgrom99, Smolin_2017, Cadoni_2019}.

As the internal acceleration of DF44 is significantly below $a_0$, the choice of interpolation function has a negligible effect on our results. Here, we use the standard function $\mu (x) = x/\sqrt{1+x^2}$. With this function, the MOND acceleration $g$ is related to the Newtonian acceleration $g_N$ as follows \citep{Milgrom83b}:  %Haghi_2016
\begin{equation}
	g ~=~ g_N \sqrt{\frac{1}{2}+\frac{1}{2} \sqrt{1+4 \left( \frac{a_0}{g_N} \right)^2}} \, ,
\end{equation}
where $g_N \equiv GM/r^2$ at some 3D radius $r$ within which the enclosed baryonic mass $M \left( r \right) = L_I \left( r \right) \times M/L_I$. This is determined by combining photometric observations of $L_I$ with the chosen or fitted stellar mass-to-light ratio $M/L_I$.

To assess how DF44 should behave in MOND, we can apply the deep-MOND virial relation \citep{Milgrom_1995} to estimate that its globally averaged one-dimensional velocity dispersion should be $\sqrt[4]{4GMa_0/81} = 22$ km/s for an I-band luminosity of $3 \times 10^8 L_{I\odot}$ \citep{vanDokkum+19} and a stellar mass to light ratio of 1 Solar unit in this band (Sec. \ref{SPS}). This is rather similar to their reported velocity dispersions, suggesting that MOND may be consistent with DF44.

%The MOG modified acceleration law for a point source in weak gravitational fields and low velocities, derived from the MOG Lagrangian and field equations and the test particle equation of motion

\subsection{Jeans analysis}  %For 

In order to find $\sigma_{los}$ in a non-rotating spherically symmetric system, we use the Jeans equation \citep{BT08}:
\begin{equation}
  \frac{d(\nu(r) \sigma_{r}^2(r))}{dr}+\frac{2\nu(r)}{r} \beta(r) \sigma_{r}^2(r) ~=~ -\nu(r) \frac{d\Phi}{dr},
\end{equation}
where $r$, $\nu(r)$, $ \beta(r)$, $\sigma_{r} \left( r \right) $ and $\Phi(r) $  are the radial distance from the center of the galaxy, the spatial number density of stars, the velocity anisotropy, the radial velocity dispersion as a function of radial distance and the gravitational potential, respectively. In general, $\beta $ could be zero (i.e. isotropic velocity dispersion tensor), constant or a function of $r$.

%it cab be calculated:
%\begin{equation}
%\label{e10}
% \frac{d \sigma_{r}^2(r)}{dr} + \frac{A(r) \sigma_{r}^2(r)}{r}=-a(r),
 % \end{equation}
%where, $A(r)=2\beta(r)+\gamma(r) $ and $ \gamma(r)=\mathrm{d} \:\ln\nu(r)/\mathrm{d}\: \ln r $. 

The observable $\sigma_{los}$ is given by
\begin{equation}
\label{e9}
\sigma^2_{\textit{los}}(R) ~=~ \frac{\int_{0}^{\infty} [y^2+(1-\beta(r))R^2]r^{-2} \sigma_{r}^2(y) \nu (y) dy}{\int_{0}^{\infty} \nu (y) dy},
\end{equation}
where $R$ and $r$ are the 2D and 3D distances from the center of the galaxy, respectively, and  $ y \equiv \sqrt{r^2-R^2} $. 

%Substituting $y$ into Eq. \ref{e9}, $\sigma_{\textit{los}}(R)$ can be written as \citep[][eq. 6]{Lokas_2005}
%\begin{equation}
%\label{e12}
 % \sigma^2_{\textit{los}}(R)\!=\!\frac{\int_{R}^\infty (r^2\!-\!\beta(r)R^2)\sigma_{r}^2(r) \nu(r)/r\sqrt{r^2\!-\!R^2}\: dr}{\int_{R}^\infty r \nu(r)/\sqrt{r^2-R^2} \:dr }.
%\end{equation}

For the density distribution, we use a Plummer model \citep{Plummer_1911}:
\begin{equation}
\nu (r) ~=~ \left( \frac{3M}{4 \pi a^3_p} \right) \left( 1+\frac{r^2}{a^2_p} \right)^{-\frac{5}{2}} \, ,
\end{equation}
where $M$ is the total mass of the galaxy and $a_p$ is its Plummer radius, which is $\approx 1.3 \times$ smaller than the 3D half-light radius $r_h$ \citep{Wolf10}. In this study, we assume $r_h = 3.5$ kpc and a total luminosity of $L_I = 3 \times 10^ 8 L_{I\odot}$ \citep{vanDokkum16, vanDokkum+19}.

%The los-velocity dispersion which is the directly measurable dSph property as a function of projected distance from the centre of galaxy is given by

\begin{table}
%\begin{center}
\centering
\begin{tabular}{cccccc}
\hline
\hline
 Model &$M_{dyn}/L_{I}$ &$\beta$& $\alpha$ &$\mu [kpc^{-1}]$ & $\chi^2_N $ \\
&&&&&(P-value)\\
\hline
 MOG 1 & $7.4_{-1.4}^{+1.5}$  &$-0.1_{-0.4}^{+0.2} $&$13$ (fixed) &$0.15$ (fixed)&$ 0.43(0.15)$ \\

%  &$M_{dyn}/L_{I}$ &$\beta$& $alpha$ &$\mu [kpc^{-1}]$ & $\chi^2_N $ \\
\hline
MOG 2& $3$ (fixed) &$0.0_{-0.25}^{+0.20} $&$94_{-19}^{+20}$ & $0.09_{-0.01}^{+0.01}$&$ 0.47(0.25)$ \\

%   &$M_{dyn}/L_{I}$ &$\beta$& $alpha$ &$\mu [kpc^{-1}]$ & $\chi^2_N $ \\
\hline
   MOG 3 & $1$ (fixed)  &$0.0_{-0.25}^{+0.20} $&$109_{-20}^{+21}$ &$0.18_{-0.03}^{+0.02} $ &$ 0.52(0.29)$ \\
%  MOND  & $M_{dyn}/L_{I}$ & $\beta$ & & &$\chi^2_N $ \\
\hline
 \hline
MOND & $3.6_{-1.2}^{+1.6}$ & $-0.5_{-1.6}^{+0.4}$& -- & -- &$ 0.58(0.25) $ \\

 \hline
 \hline
\end{tabular}
%\end{center}
\caption{The results of fitting MOND and MOG models to the observational line-of-sight velocity dispersion profile, which we extracted from \citet{vanDokkum+19}.}
\label{t1}
\end{table}

\begin{figure*}
\includegraphics[width=60mm, height= 50mm]{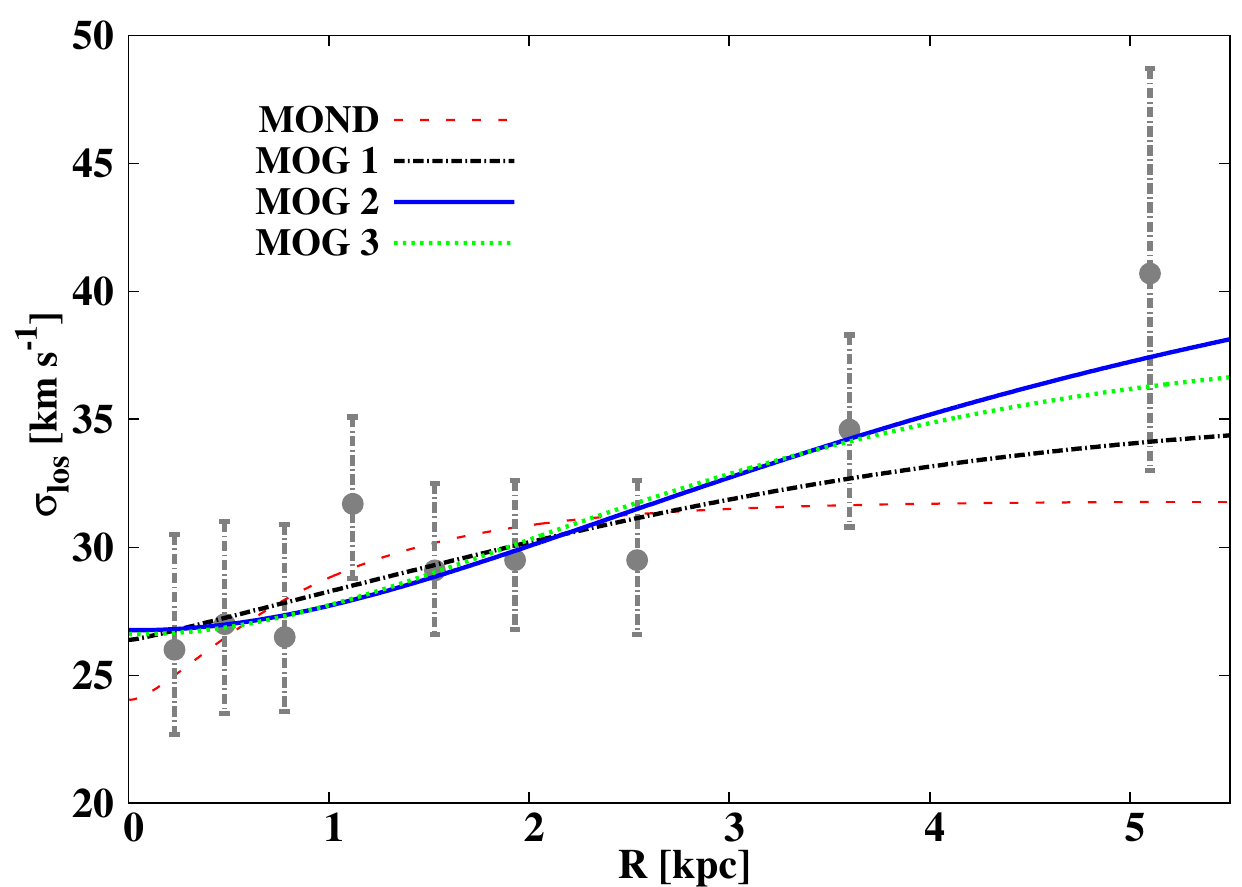}
\includegraphics[width=60mm, height= 50mm]{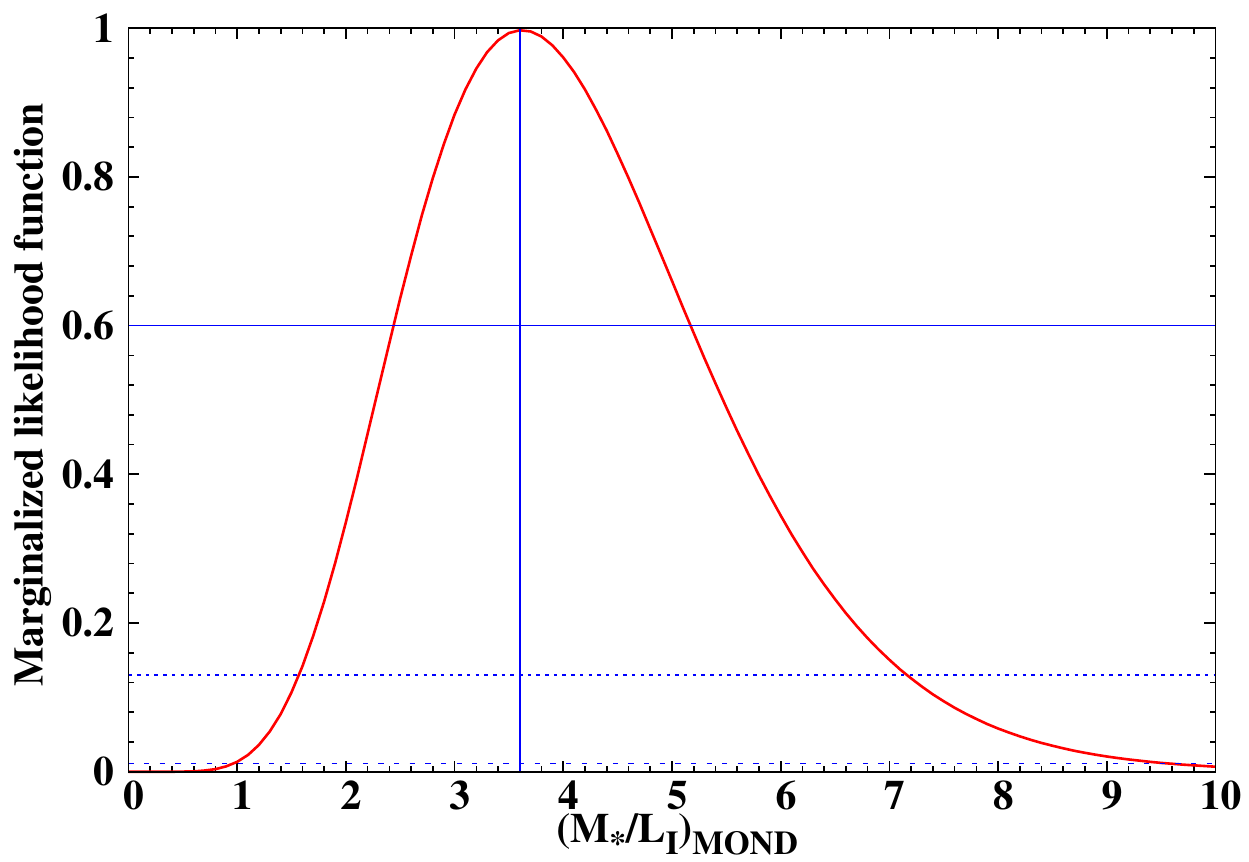}
\includegraphics[width=60mm, height= 50mm]{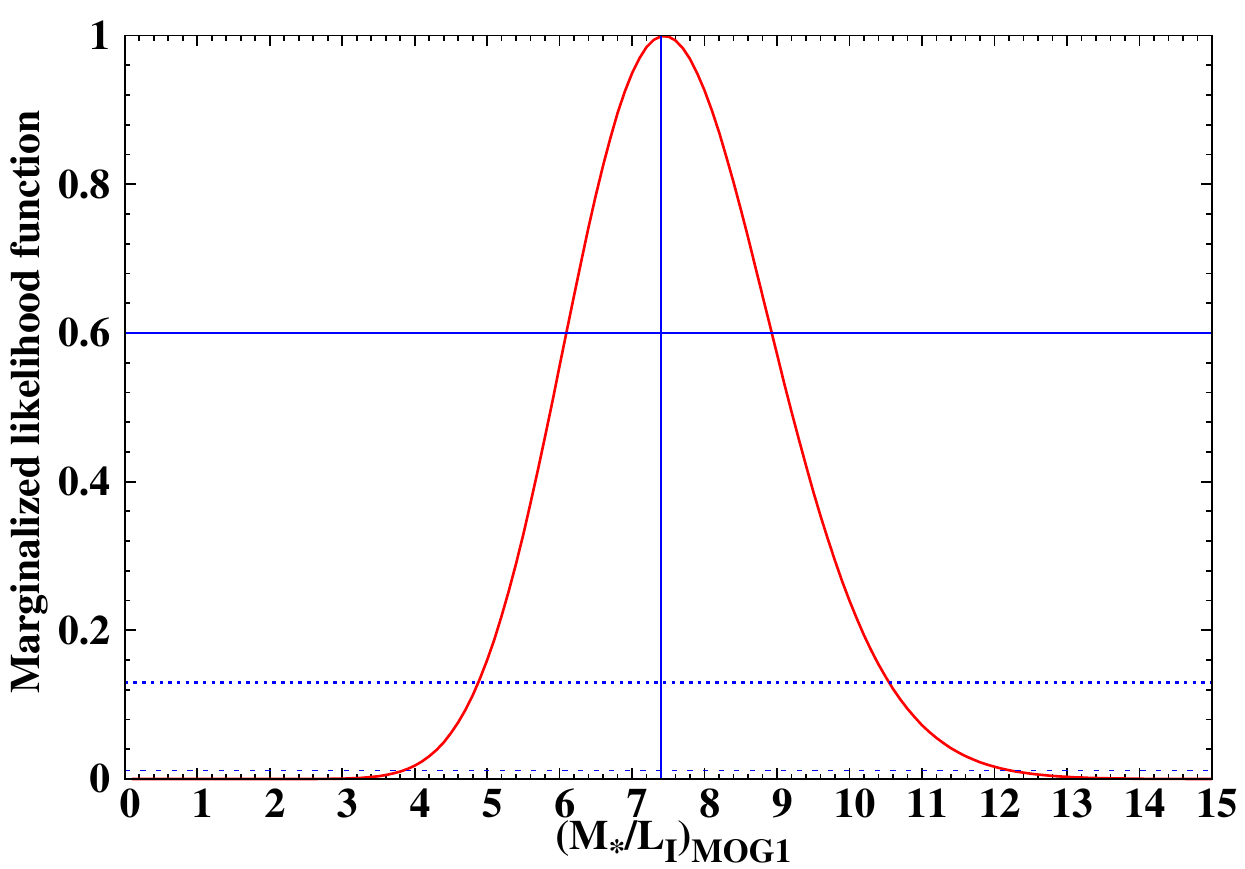}
\caption{Left: The best fits to the DF44 $\sigma_{\textit{los}}(R)$ profile \citep{vanDokkum+19} obtained from the Jeans equation in MOND and MOG models. The $x$-axis is the sky-projected distance. Middle: the marginalized 68\% (solid line), 95\% (dotted line) and 99\% (dashed line) confidence levels on $M_{dyn}/L_I$ for the MOND model. Right: same as the middle panel but for the MOG1 model.}
\label{f1}
% \end{center}
\end{figure*}

\section{Results} \label{Results}

By adjusting the free parameters of a model, the fit to the data can be improved or worsened. We quantify the goodness of each fit using the reduced $\chi^2$ statistic, which is defined as
\begin{equation}
 \chi_N^2 ~=~ \frac{1}{(N-P)}\sum_{i=1}^{N}\frac{(\sigma_{los,theory}^i-\sigma_{los,obs}^i)^2}{\sigma_{i}^2 }\label{chi} \, ,
 \label{chi_sq}
\end{equation}
where $\sigma_i$ is the uncertainty of observed data point $i$, $N=9$ is the number of data points and $P$ is the number of degrees of freedom in the model.

 \subsection{MOG fits} \label{MOG_fits}

By applying the formalism described in Sec. \ref{Basics} to the observed data of DF44, we calculate its $\sigma_{los}$ profile. Given that DF44 has $L_I = 3\times 10^8 L_{I\odot}$, its baryonic mass is within the range of galaxies in the THINGS catalog. Thus, in the first step (MOG1), we fix the MOG parameters to $\alpha_{RC}=13$ and $\mu_{RC}=0.15 kpc^{-1}$, which are upper limits inferred from fitting MOG to galaxy RCs \citep{Moffat13}. 

We allow variation in the $M_{dyn}/L_I$ ratio and anisotropy parameter $\beta$. As shown in Table \ref{t1}, although a good fit is achieved with reduced $\chi^2=0.43$, the inferred $M_{dyn}/L_I$ of $7.4^{+1.5}_{-1.4}$ for model MOG1 is larger than expected from SPS models (Fig. \ref{f1}). We also derive the uncertainty on the $M_{dyn}/L_I$ ratio. The marginalized 1$\sigma$, 2$\sigma$ and 3$\sigma$ confidence intervals on $M_{dyn}/L_I$ are indicated in the right-hand panel of Fig. \ref{f1} as horizontal lines.

In order to produce higher values of $\sigma_{los}$ with a lower $M_{dyn}/L_I$, one needs to assign larger values of $\alpha$ and $\mu$ compared to $\alpha_{RC}$ and $\mu_{RC}$. Given that the above $\alpha_{RC}$ and $\mu_{RC}$ are upper limits, we conclude that the best-fitting parameters are inconsistent with limits imposed by rotation curve analyses. Indeed, it has already been shown that these parameters should not be taken as universal constants but should be considered as mass-dependent \citep{Moffat06, Haghi10, Haghi2016}. For example, fits to $\sigma_{los}$ data of the MW dSph galaxies yield average values of $\alpha_{dSph}=221\pm 112$ and $\mu_{dSph}=0.41 \pm 0.35$ kpc$^{-1}$ \citep{Haghi2016}.

Therefore, in the second step, we let these two parameters vary and fit the velocity dispersion data with a fixed $M/L_I=3$ (model MOG2) or 1 (model MOG3) Solar units. Fig. \ref{f1} shows that the MOG theory can successfully reproduce the observed data with acceptable values of $\chi ^2$. In these cases, MOG prefers an isotropic velocity dispersion tensor of DF44 (i.e. $\beta \approx 0$).

As shown in Table \ref{t1}, the best-fitting values of $94\leq\alpha\leq109$ and $0.09\leq\mu\leq0.18$ kpc$^{-1}$ are consistent with the results for dSph galaxies \citep{Haghi2016}. However, the best-fitting $\alpha$ values are larger than those inferred from rotation curve analysis of spiral galaxies, supporting the hypothesis that the MOG parameters are mass-dependent. This is consistent with the trend claimed by \citet{Brownstein2007} that the best-fitting values of $\alpha$ and $\mu$ decrease when moving from low-mass systems (e.g. dwarf galaxies) to massive systems (e.g. dwarf and normal X-ray clusters). However, MOG does not offer a physical explanation for such a dependency.

\subsection{MOND fit} \label{MOND_fit}

By solving the Jeans equation, we calculate $\sigma_{los} \left( R \right)$ in the MOND framework. The best-fitting values of the mass-to-light ratio and anisotropy parameter along with the minimum $\chi ^2$ value are given in Table \ref{t1}. The MOND fit to the $\sigma_{los}$ profile is shown in Fig. \ref{f1} as a red dashed line. The reduced $\chi^2$ of $0.58 $ is quite good, as can also be seen in Fig. \ref{f1} $-$ the best-fitting profile passes through 8 out of 9 data points within their 1$\sigma$ uncertainties, while the prediction for the outermost point is only just outside its 1$\sigma$ range. In this model, the anisotropy is negative ($\beta = -0.5$) i.e. the tangential velocity dispersion exceeds the radial one. In addition, to be comfortably explained in MOND, DF44 should have a $M_{dyn}/L_I$ ratio of around 3.6 Solar units. Our result is in good agreement with those of \citet{Bilek_2019}, who found the tangentially anisotropic model with $\beta = -0.5$ and $M_{dyn}/L_I = 3.9$ provides a reasonable fit to the data (see their fig 2b).

% These values are significantly larger than $\alpha_{RC}$ and $\mu_{RC}$ inferred from the best fits to the rotation curve of spiral galaxies.  

%The implied M∗/L ratios are given in Tables 1 and 2. Accounting for ae = 0 leads to a higher estimation of M∗/L. The uncertainties on the best-fitted values of M∗/L has been derived from the 68 per cent confidence level.  

%The 1σ and 2σ confidence regions for the best-fitting parameters to the observed RCs of 18 dwarf and spiral galaxies listed in Tables 1 and 2 are shown in Fig. 3. The central points correspond to the best-fitting values

As can be seen in Fig. \ref{f1}, the $M_{dyn}/L_I$ ratio of $3.6_{-1.2}^{+1.6}$ inferred from our MOND dynamical model of DF44 (shown as horizontal lines) can be acceptable in the context of the IGIMF theory if the star formation occurred very rapidly (within the first 50-80 Myr) and shut off thereafter (Fig. \ref{M_L_DF44}).

At 3$\sigma$, our MOND dynamical modeling is consistent with $M_*/L_I = 1.0$ (middle panel of Fig. \ref{f1}). The significant uncertainty in this is a consequence of the fact that $\sigma \propto \sqrt[4]{M}$ in the deep-MOND limit, allowing large changes in $M_*/L_I$ to have a relatively modest impact on the observed kinematics. A mass-to-light ratio of $M_*/L_I = 1.0$ arises in our IGIMF model for $\Delta t = 1$ Gyr (Fig. \ref{M_L_DF44}). This is longer than DF44's dynamical timescale of $t_{dyn} = r_h/\sigma \approx 66$ Myr if $r_h = 3.5$ kpc and the 3D velocity dispersion $\sigma$ is $\sqrt{3} \times$ the observed $\sigma_{los}$ of 30 km/s.

A short star formation timescale, being consistent with  MOND,  might also naturally explain the large size of DF44, since gas expelled by massive stars would cause it to expand \citep{Wu_2018}. This result is in line with the suggestion by \citet{vanDokkum+18a} that UDGs must have had extremely high gas densities at the time of their formation and perhaps feedback from an intense, compact star burst that created them caused both the cessation of star formation and their expansion to become UDGs.

In the context of the IGIMF theory, the slope of the gwIMF above 1 $M_\odot$ varies with the SFR. A higher SFR leads to the gwIMF being more top-heavy, while the slope of the gwIMF does not change for $m < 1 M_\odot$ and remains the same as the canonical IMF (but see \citealt{Jerabkova18}). Adopting an age of 10 Gyr and $\Delta t < 500 Myr$, only stars with $m < 1 M_\odot$ remain alive on the main sequence and contribute to the total luminosity. Stars with higher masses have already evolved and turned into remnants, so they contribute to the total mass but not to the luminosity. Since we fix the $I$-band luminosity of the modeled galaxies to be the same as DF44, adopting the IGIMF instead of the canonical IMF only changes the total mass of DF44 and hence its $M_*/L_I$. DF44's suggested short star formation duration implies that its color will be independent of $\Delta t$ as long as this is much less than its age.%(combined with DF44's age and short star formation duration) only  the total mass and hence the mass-to-light ratio of galaxy will change and its luminosity and colors in different bands will be unchanged.

%The timescale required in MOG is about 30 Myr, which is very short compared to the dynamical timescale of t_dyn = r_h/sigma_los = 65 Myr. However, the MOND model is consistent at 3 sigma with a dynamical M/L of 1, which can be attained in the framework of the canonical IMF or the IGIMF with dt = 500 Myr, which we consider to be reasonable given t_dyn.

%

%The 1σand 2σconfidence regions for the best-fitting parameters tothe observed RCs of 18 dwarf and spiral galaxies listed in Tables1and2are shown in Fig.3. 

\subsection{Consistency with observations}
\label{Section_consistency}

In this section, we show the joint constraint on the $M/L_I$ and $\Delta t$ of DF44 based on dynamical modeling and typical values of $\Delta t$ as found by \citet{Pflamm09}, respectively. We also show the locus of $\left( M/L_I, \Delta t \right)$ values consistent with stellar population modeling in the IGIMF framework (Sec. \ref{SPS}).

\begin{figure}
\includegraphics[width=80mm, height= 55mm]{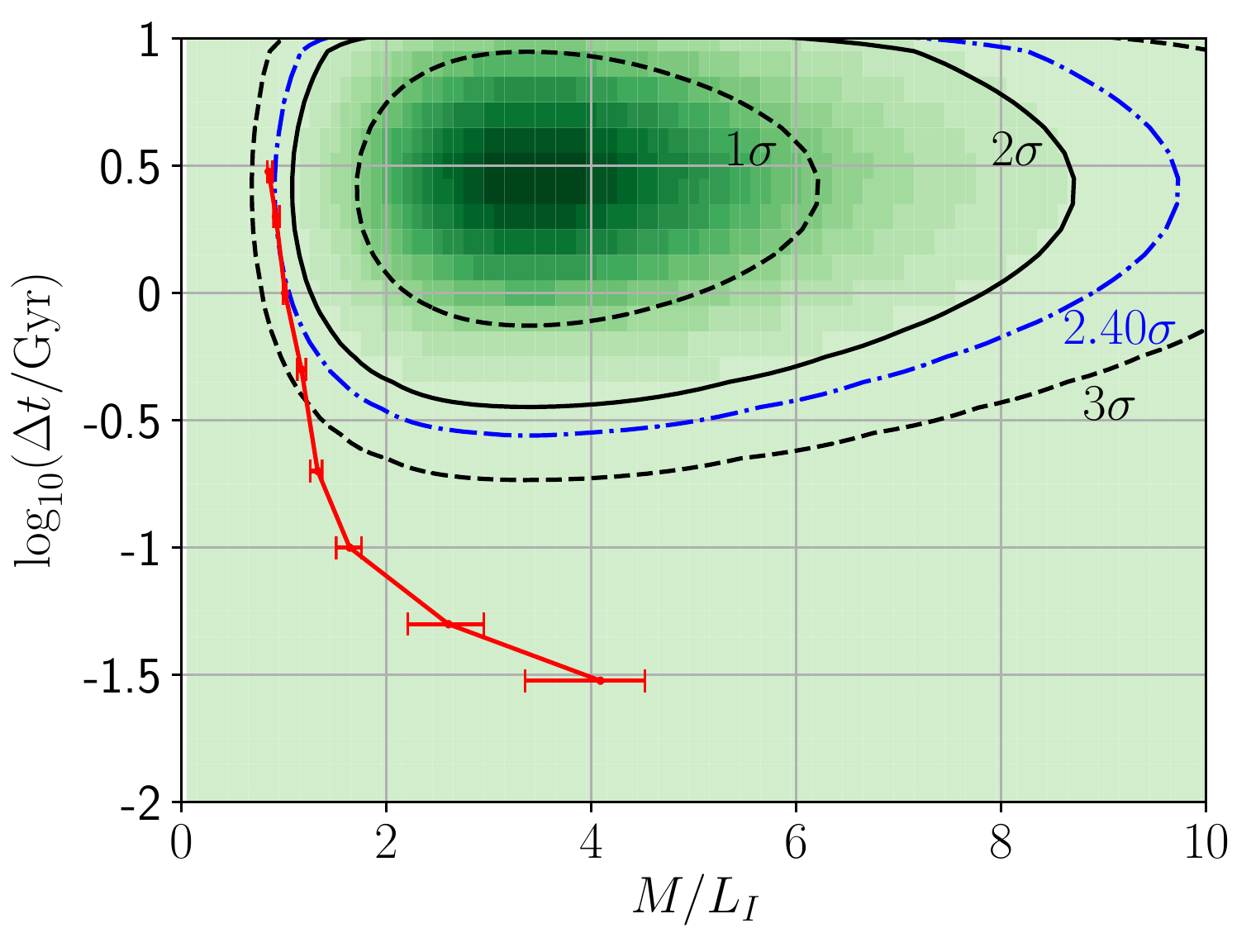}
\includegraphics[width=80mm, height= 55mm]{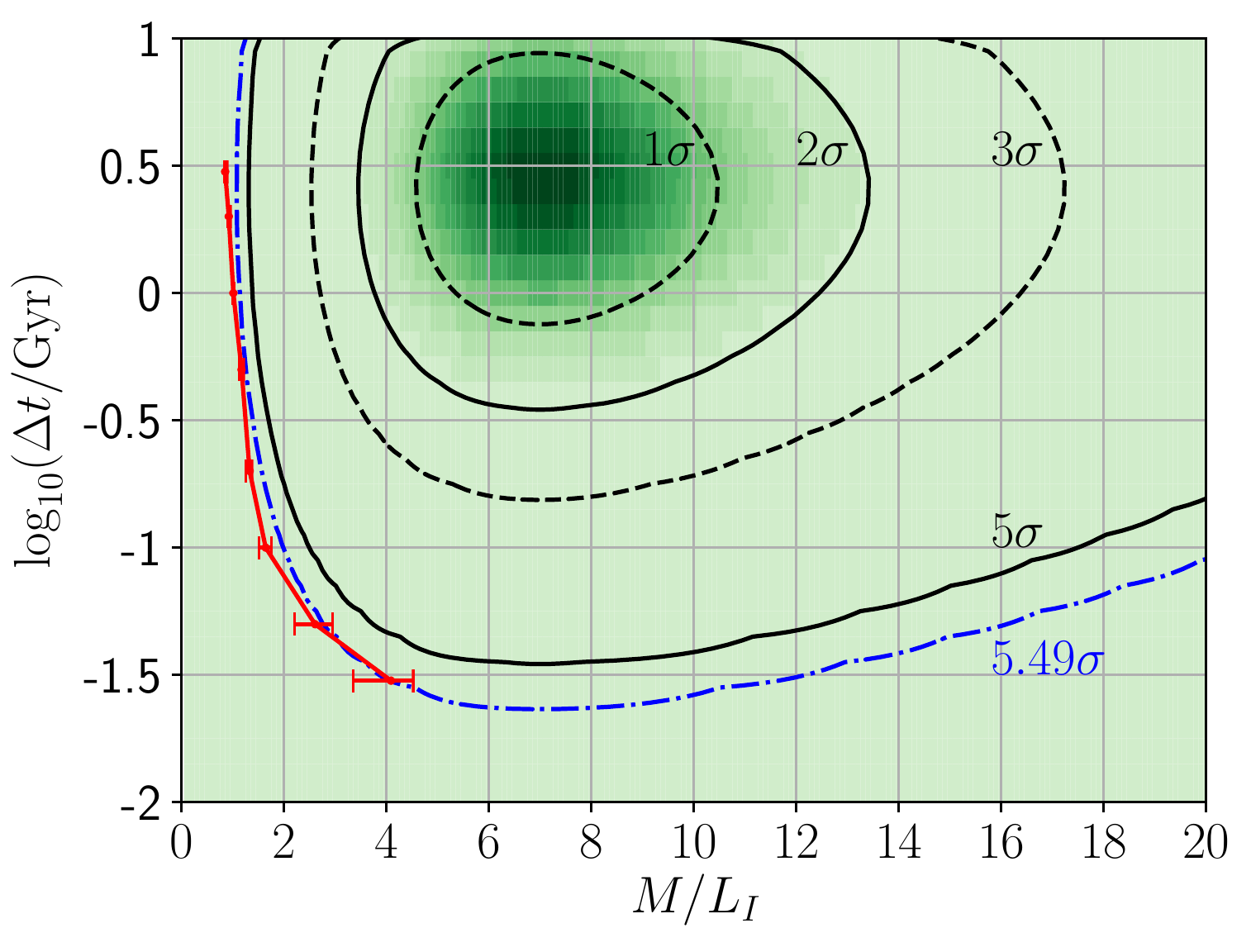}
\caption{Indicated contours of the joint probability of the stellar $M_{dyn}/L_I$ ratio and $\mathrm{log}_{10}(\Delta t/Gyr)$ for MOND (top panel) and MOG1 (bottom panel), calculated with Eq. \ref{P_joint}. In both panels, the red tracks are $M_*/L_I$ values of our SPS models in the IGIMF framework for $L_I = \left(3.0 \pm 0.6 \right) \times 10^8 L_{I\odot}$ (Sec. \ref{M_L_DF44}) while the blue contours marginally intersect a point on these tracks.}
\label{f2}
% \end{center}
\end{figure}

We begin by plotting a 2D array of probabilities as a function of $M_{dyn}/L_I$ and $\mathrm{log}_{10}(\Delta t)$ (with $\Delta t$ in Gyr). We then show the 68.3\%, 95.4\% and 99.7\% confidence levels of the probability distribution (Fig. \ref{f2}). The probability $P$ is the product of the posterior on $M_{dyn}/L_I$ from the kinematics (Eq. \ref{chi_sq}) multiplied by the likelihood of a particular star formation timescale, i.e.
\begin{eqnarray}
    P \left(M_{dyn}/L_I, \Delta t \right) = P \left( M_{dyn}/L_I \right) \times P \left( \Delta t \right) \, , \text{where} \nonumber \\
    P \left( \Delta t \right) \propto \exp \left( -\frac{\left(\mathrm{log}_{10} \Delta t - \mathrm{log}_{10}\Delta t_{expected} \right)^2}{2\sigma^2} \right)
    \label{P_joint}
\end{eqnarray}
and $\Delta t_{expected} = 2.9$ Gyr is the expected $\Delta t$ from  \citet{Pflamm09} for an assumed gas mass of $10^6 M_\odot$ (using a $10\times$ lower or higher value has a negligible effect). According to section 5 of their work, $\Delta t$ has a dispersion of $\sigma = 0.36$ dex. We impose an upper limit of $\Delta t \leq 10$ Gyr since the red colour of DF44 shows it did not form stars very recently. Since our dynamically inferred posteriors on $M_{dyn}/L_I$ (Fig. \ref{f1}) are calculated for fixed $L_I = 3\times 10^8 L_{I\odot}$, we convolve our $P \left( M/L_I \right)$ with a Gaussian in $\mathrm{log}_{10}(L_I)$ of width $0.088 = \frac{1}{2} \left( \mathrm{log}_{10}(1.2) - \mathrm{log}_{10}(0.8) \right)$, thereby accounting for a 20\% uncertainty in $L_I$ \citep[section 6.2 of][]{vanDokkum+19}.

To quantify the consistency of each model with observations, we change the confidence level until the corresponding contour just intersects one point on the SPS track of $\left(M_*/L_I, \Delta t\right)$. In this way, we find that MOND is consistent with observations at 1.66\%, which corresponds to $2.40 \sigma$ for a Gaussian distribution. However, MOG1 is consistent with observations only at the $4.07 \times 10^{-8}$ level ($5.49 \sigma$), so our results rule out MOG in its present form $-$ we are already setting its free parameters to the upper limits allowed by THINGS (Sec. \ref{MOG_fits}).

The MOND model marginally matches the data if we bear in mind that there must occasionally be $3 \sigma$ outliers from the correct theory. Given that we have accurate kinematics for $\approx 200$ galaxies and MOND works well in the vast majority of them \citep{Li_2018}, we expect to come across a few cases where the consistency is only at the percent level. Indeed, it would be somewhat unusual if MOND explained all observed galaxies within $2.4 \sigma$. However, there should be no $5.5\sigma$ events within a sample of this size, showing that presently available data on DF44 rule out the MOG model if its free parameters are universal \citep*[as claimed by its proponents,][]{Green_2018} and our other modeling assumptions are correct. If DF44 has some rotation within the plane of the sky, its self-gravity would be even stronger than we assumed, thus requiring an even larger $M_{dyn}/L_I$ and making the situation worse for MOG. The same is true if we use the canonical IMF instead of the IGIMF as the former predicts $M_*/L_I = 1$ (Fig. \ref{M_L_DF44}).

\vspace{20pt}
\section{Discussion} \label{Discussion}

\cite{vanDokkum+19} recently reported a revised stellar velocity dispersion within the effective radius for the UDG known as DF44, implying its Newtonian  $M_{dyn}/L_I \approx 26$ Solar units. They argue that the galaxy is gravitationally dominated by dark matter, in apparent contrast to the UDGs DF2 and DF4. 
%They also found that the velocity dispersion profile of DF44 cannot be fit with a standard NFW halo and an isotropic velocity distribution. Only a cored halo profile \citep{DiCintio14} fits the data with isotropic orbits. 
They claimed that the MOND-predicted velocity dispersion of DF44 is lower than the observed value, challenging alternatives to dark matter theories such as MOND. 

We took DF44 to be in equilibrium by adopting a simplified density profile without any tidal interaction. 
%This is consistent with a rising velocity dispersion profile, which is in fact a prediction of our model (Fig. \ref{f1}). 
Tidal disruption scenarios cannot easily be reconciled with the high GC counts of UDGs in Coma and their lack of obvious tidal features. The fact that DF44 appears to live in a dynamically cold environment \citep[][their section 6.1]{vanDokkum+19} can be interpreted as evidence against tidal heating by the Coma cluster. However, some tidal heating could increase the velocity dispersion in the outer parts of DF44, easing the tension with MOND \citep{Bilek_2019}.

In addition to neglecting tides, our MOND calculations assume no external field effect (EFE; see \citealt{Famaey12} for a review and discussion of the MOND-unique EFE; see also \citealt{WK15} for an accessible description). Since the internal acceleration is $g_{int} \approx \sqrt{GMa_0}/r_h = 0.24 \, a_0$ for $M_*/L_I = 1$, the EFE would not significantly affect our results as long as the external field (EF) is much weaker than this. The MOND dynamical mass of the Coma cluster is $4.6 \times 10^{14} M_\odot$ \citep{Sanders_2003}. This implies DF44 is $\gg 3.0$ Mpc from the center of the Coma cluster, in which case DF44 is not in the cluster. This is entirely consistent with its sky-projected separation of $\approx 1.7$ Mpc \citep[][their section 3.1]{Bilek_2019} and the observation that two other galaxies, DF42 and DFX2, have a radial velocity within 100 km/s of the value for DF44, suggesting these are part of a different group which is dynamically much colder than the Coma cluster \citep[][their section 6.1]{vanDokkum+19}. For a MW-like galaxy with $M = 10^{11} M_\odot$ to provide an EF comparable to $g_{int}$, the separation with DF44 would need to be $\la 24$ kpc. There is no such galaxy so close to DF44, strengthening confidence in our assumption that it can be treated as isolated.

So far, we have assumed that DF44 is at a downrange distance of $D = 100$ Mpc based on its redshift. However, it may have a peculiar velocity of a few hundred km/s. The MOND-predicted $\sigma_{los} \propto \sqrt{D}$ for fixed $M/L_I$ and apparent magnitude \citep{Kroupa_2018}. Since $\sigma_{los} \propto \sqrt[4]{M}$ in the deep-MOND limit \citep{Milgrom_1995}, the MOND $M_{dyn}/L_I \propto D^{-2}$. Thus, placing DF44 just 10\% further away reduces the best-fitting $M_{dyn}/L_I$ to 3.0, improving the agreement with SPS expectations. Its observed velocity dispersion could also have been overestimated slightly.

Our analysis prefers a relatively short formation time scale for DF44. This can occur if it is a tidal dwarf galaxy (TDG) and lost its gas reservoir by ram pressure or tidal stripping i.e. if it was quenched \citep{Weisz_2015}. The temporal evolution of the quenched fraction can also provide clues as to how quickly dwarf galaxies can undergo quenching. For example, in the lowest-mass systems, nearly 40\% already appear to be quenched by $\approx 12$ Gyr ago, which implies a quenching timescale of 1-2 Gyr \citep{Weisz_2015}. Therefore, DF44 may well have been quenched within 1 Gyr of its formation.

Once there is no further gas supply, the available gas is consumed on a dynamical time-scale. Pressure supported systems do indeed form within a few dynamical times, as is nicely evident in elliptical galaxies $-$ see the downsizing results by \citet{Thomas_1999} and \citet{Recchi09}. Since DF44 is pressure-supported \citep{vanDokkum16}, it may well have a somewhat shorter $\Delta t$ than estimated by \citet{Pflamm09} since their work only considered late-type galaxies.

\section{Conclusion} \label{Conclusion}

We constructed SPS models of DF44 in the IGIMF framework to estimate its stellar $M_*/L_I$ ratio independently of the assumed gravity law. We then used spatially-resolved kinematic data of DF44 to infer its dynamical $M_{dyn}/L_I$ in the framework of MOG and MOND, which are alternative approaches to the cold dark matter hypothesis. The main conclusions of these calculations are as follows:

\begin{itemize}
    \item First, we used MOG with the upper limits $\alpha_{RC}=13$ and $\mu_{RC}=0.15 kpc^{-1}$ inferred from fits to galaxy rotation curves \citep{Moffat13}. The only free parameter was the stellar $M_*/L_I$ ratio. We found that these models provide a reasonably good fit, but the required high near-infrared $M_{dyn}/L_I$ ratio of $\approx 7.4$ Solar units is completely inconsistent with stellar population synthesis modeling for any plausible star formation duration.

     \item Then, we let $\alpha$ and $\mu$ vary as free parameters and fitted the velocity dispersion data assuming $M_*/L_I$ = 1 and 3 Solar units. The best-fitting values of $\alpha$ and $\mu$ are larger than $\alpha_{RC}$ and $\mu_{RC}$, supporting the hypothesis that the MOG parameters are mass-dependent. They are compatible with the average values of $\alpha=221 \pm 112$ and $\mu=0.41 \pm 0.35 kpc^{-1}$ obtained from fitting MOG to $\sigma_{los}$ data of the MW dSph galaxies \citep{Haghi2016}. These values are significantly larger than $\alpha_{RC}$ and $\mu_{RC}$ inferred from fits to the rotation curves of spiral galaxies \citep{Moffat13}.

     \item We calculated the $\sigma_{los}$ profile of DF44 in MOND and found that the best fitting model with constant orbital anisotropy has $\beta = -0.5_{-1.6}^{+0.4}$ and a stellar mass-to-light ratio of $M_{dyn}/L_I=3.6_{-1.2}^{+1.6}$ Solar units. Obtaining $M_*/L_I = 3.6$ requires a star formation duration $\Delta t \approx 40$ Myr. However, the $3 \sigma$ lower limit on $M_{dyn}/L_I$ is 1.0, consistent with the canonical IMF and expectations from SPS modeling in the IGIMF context for $\Delta t < 1$ Gyr (Fig. \ref{M_L_DF44}).
 
     \item By considering the joint constraints on $M/L_I$ and $\Delta t$ from dynamical modelling and observations of other galaxies, respectively, we showed that MOND is consistent with our $\Delta t = 2$ Gyr star formation model of DF44 at $2.40 \sigma$ (1.66\% confidence). The same approach shows that MOG is ruled out at $5.49 \sigma$ since it has only a $4.07 \times 10^{-8}$ chance of explaining the observations even if $\alpha$ and $\mu$ are set to the upper limits inferred from rotation curve fitting. \newline
\end{itemize}

\vspace{20pt}
\section*{Acknowledgements}

AHZ and IB are supported by Alexander von Humboldt postdoctoral research fellowships. The authors wish to thank the reviewer for very detailed comments which significantly improved this manuscript.

\vspace{20pt}
\bibliographystyle{aasjournal}
\bibliography{DF44_bbl}
\label{lastpage}
\end{document}